\documentclass{article}
\input{tcilatex}

\begin{document}

\title{New Glance at the Experimental Data for Low Lying Collective Excited
States.}
\author{Vladimir \ P. \ Garistov \\
%EndAName
Institute \ for \ Nuclear Research and \ Nuclear Energy}
\maketitle

\begin{abstract}
Recently \ the classification of \ low-lying \ excited $0^{+}$ states \ in
even even deformed nuclei \ has been done. The available experimental data
were \ represented \ as \ the \ energies parabolic distributed \ by number
of \ monopole \ excitations. With \ other \ words \ each $0^{+}$ state now
is determined as the collective state with the corresponding number of \
monopole type phonons \ $n$ .\ In this short remark \ we discuss whether \
the experimental data for low-lying excited states possessing not equal to
zero spins can also be described with \ parabolic distribution \ function
depending on integer classification parameter \ and find any vindication of
the connection between \ this integer parameter and the \ number \ of
collective excitations building the \ corresponding state. \ 
\end{abstract}

In our recent investigations \cite{1} of the yrast lines in even-even
deformed nuclei we obtained that the energies of these lines can be
described with great accuracy even if we use the simple rigid rotor model
but if we consider yrast line be built with several number of crossing rigid
rotor bands and if we make the bands heads be responsible for the behavior
of \ the rotational bands. With other words we make the band head be
responsible for the value of the moment of inertia of the nucleus staying in
corresponding excited \ state.

We build the \ positive \ parity lines with crossing of several number of
rotational $\beta $ - bands starting from different excited $0^{+}$states
that \ we consider as their heads. To \ understand \ the peculiarities of
different excited $0^{+}$states we analyzed a \ great amount of experimental
data for low \ lying excited $0^{+}$states in even-even nuclei. \ \ 

We represent the available experimental data in the form of the energies of
the $0^{+}$excited states distributed by positive integer parameter and \
determine this classification parameter in the \ way giving \ us information
about collective structure peculiarities of these states.

\qquad To specify the distribution function let us consider the monopole
part of \ collective Hamiltonian for single level approach written in terms
of boson creation and annihilation operators\ \ $R_{+}$ , $R_{-}$ and $R_{0}$%
\begin{equation}
H=\mathbf{\alpha }R_{+}^{j}R_{-}^{j}+\mathbf{\beta }R_{0}^{j}R_{0}^{j}+\frac{%
\mathbf{\beta }\Omega ^{j}}{2}R_{0}^{j},  \label{HamR}
\end{equation}

\bigskip constructed with the pairs of fermion operators $a^{\dagger }$ and $%
a$.%
\begin{equation}
\begin{tabular}{l}
$R_{+}^{j}={\frac{1}{2}}\sum\limits_{m}(-1)^{j-m}\alpha _{jm}^{\dagger
}\alpha _{j-m}^{\dagger }\;$, \\ 
$R_{-}^{j}={\frac{1}{2}}\sum\limits_{m}(-1)^{j-m}\alpha _{j-m}\alpha
_{jm}\;, $ \\ 
$R_{0}^{j}={\frac{1}{4}}\sum\limits_{m}(\alpha _{jm}^{\dagger }\alpha
_{jm}-\alpha _{j-m}\alpha _{j-m}^{\dagger })\;.$ \\ 
$\left[ R_{0}^{j},R_{\pm }^{j}\right] =\pm R_{\pm }^{j},\ \ \ \ \ \ \ \left[
R_{+}^{j},R_{-}^{j}\right] =2R_{0}^{j}$%
\end{tabular}
\label{ROPER}
\end{equation}

\ 

Further applying the Holstein-Primakoff \cite{2} transformation to the
operators $R_{+}$ , $R_{-}$ and $R_{0}$

\bigskip\ \ \ \ \ \ $\ \ \ \ \ \ \ 
\begin{array}{ccc}
R_{-}=\sqrt{2\Omega -b^{\dagger }b}\;b\text{; \ \ \ } & R_{+}=b^{\dagger }%
\sqrt{2\Omega -b^{\dagger }b}\text{; \ \ \ } & R_{0}=b^{+}b-\Omega .%
\end{array}%
$%
\[
\left[ b,b^{\dagger }\right] =1\,,\;\left[ b,b\right] =\left[ b^{\dagger
},b^{\dagger }\right] =0.\ \ \ \ 
\]

the initial Hamiltonian (\ref{HamR}) \ written in terms of pure bosons has
the form:

\begin{equation}
H=Ab^{\dagger }b-Bb^{\dagger }bb^{\dagger }b.  \label{Hampure}
\end{equation}

\[
A=\alpha (2\Omega +1)-\beta \Omega ,\,\,B=\ \alpha -\beta . 
\]%
\[
\ \ \left\vert n\right\rangle =\frac{1}{\sqrt{n!}}(b^{+})^{n}\left\vert
0\right\rangle ,where\ \ b\left\vert 0\right\rangle =0\;\ 
\]

Thus the energy spectrum produced by Hamiltonian (\ref{Hampure}) \ is the
parabolic function of the number of monopole bosons $n\ \ \ \ \ \ \ \ \ \ \
\ \ \ \ \ \ \ \ \ \ \ \ $%
\begin{equation}
E_{n}=An-Bn^{2}+C  \label{E(n)}
\end{equation}

This is the form we apply in our new representation of the experimental data
of the low lying excited $0^{+}$ - states. Some of the \ distributions of
the experimental energies of the excited \ $0^{+}$ \ states \ plotted \ \ 

\bigskip using (\ref{E(n)}) \ are \ shown in\textbf{\ Figure} 1.

This parabolic \ distribution (\ref{E(n)}) \ \ reproduces \ with a \ great \
accuracy experimental \ values \ of \ low lying $0^{+}$ \ excited states
energies. Similarly, \ very nice \ agreement was obtained for all available
experimental \ data of \ low lying $0^{+}$ \ excited states \ in a \ large
region of the even-even nuclei. In \textbf{Figure 2}. we show the
description of the positive parity yrast line experimental data with two
crossing rotational $\beta $ bands. Along with the comparison with
experiment there are shown the numbers of bosons $n$ for the bans heads.

Of course it is straightforward now to see whether the low lying excited
states having different from zero spin can be also represented in the same
form of the energies distributed by parabolic type function and can we
connect the new classification parameter as a measure of collectivity
determining each low lying state.

For this purpose let us shortly \ remind \ the Interacting Vector \ Boson
Model \ (IVBM) \ \ developed some \ years ago\ by A. Georgieva, \ P. Raychev
and R. \ Roussev\ \cite{3}.

\qquad IVBM is based on the introduction of two kinds \ of vector bosons
(called $p$- and $n$bosons), that \textquotedblright built
up\textquotedblright\ the collective excitations in the nuclear system. The
creation operators $u_{m}^{+}(\alpha )$ of these bosons are assumed to be $%
SO(3)$-vectors and they transform according to two independent fundamental
representations $(1,0)$ of the group $SU(3).$ The annihilation operators $%
u_{m}(\alpha )=(u_{m}^{+}(\alpha ))^{\dagger }$ transform according to the
conjugate representations $(0,1).$ These bosons form a \textquotedblright
pseudospin\textquotedblright\ doublet of the group $U(2)$ and differ in
their \textquotedblright pseudospin\textquotedblright\ projection $\alpha
=\pm \frac{1}{2}.$ \ The introduction of this additional degree of freedom
leads to the extension of the $SU(3)$ symmetry to $U(6)$ so that the two
kind of \ bosons $u_{m}^{+}(\alpha =\pm \frac{1}{2})$ transform according to
the fundamental representation $[1]_{6}$ of the group $U(6)$. The bilinear
products of the creation and annihilation operators of the two vector bosons
generate the noncompact symplectic group $Sp(12,R)$ \cite{3}: 
\[
F_{M}^{L}(\alpha ,\beta )=_{k,m}^{\sum }C_{1k1m}^{LM}u_{k}^{+}(\alpha
)u_{m}^{+}(\beta ), 
\]

\[
G_{M}^{L}(\alpha ,\beta )=_{k,m}^{\sum }C_{1k1m}^{LM}u_{k}(\alpha
)u_{m}(\beta ), 
\]

\begin{equation}
A_{M}^{L}(\alpha ,\beta )=_{k,m}^{\sum }C_{1k1m}^{LM}u_{k}^{+}(\alpha
)u_{m}(\beta ),  \label{generators}
\end{equation}%
where $C_{1k1m}^{LM}$ are the usual Clebsh-Gordon coefficients and $L$ and $%
M $ define the transformational properties of (\ref{generators}) under
rotations.

We consider $Sp(12,R)$ to be the group of the dynamical symmetry of the
model \cite{3}. Hence the most general one- and two-body Hamiltonian can be
expressed in terms of its generators . Using commutation relations between $%
F_{M}^{L}(\alpha ,\beta )$ and $G_{M}^{L}(\alpha ,\beta )$, the number of
bosons preserving Hamiltonian can be expressed only in terms of operators $%
A_{M}^{L}(\alpha ,\beta )$: 
\begin{equation}
H=\sum_{\alpha ,\beta }h_{0}(\alpha ,\beta )A^{0}(\alpha ,\beta
)+\sum_{M,L}\sum_{\alpha \beta \gamma \delta }(-1)^{M}V^{L}(\alpha \beta
;\gamma \delta )A_{M}^{L}(\alpha ,\gamma )A_{-M}^{L}(\beta ,\delta ),
\label{genham}
\end{equation}%
where $h_{0}(\alpha ,\beta )$ and $V^{L}(\alpha \beta ;\gamma \delta )$ are
phenomenological constants.

Being a noncompact group, the representations of $Sp(12,R)$ are of infinite
dimension, which makes it rather difficult to diagonalize the most general
Hamiltonian. The operators $A_{M}^{L}(\alpha ,\beta )$ generate the maximal
compact subgroup of $Sp(12,R)$, namely the group $U(6)$: 
\[
Sp(12,R)\supset U(6) 
\]%
So the even and odd unitary irreducible representations /UIR/ of $Sp(12,R)$
split into a countless number of symmetric UIR of $U(6)$ of the type $%
[N,0,0,0,0,0]=[N]_{6}$, where $N=0,2,4,...$ for the even one and $%
N=1,3,5,... $ for the odd one \ \cite{3}. Therefore the \textit{complete}
spectrum of the system can be calculated only trough the diagonalization of
the Hamiltonian in the subspaces of \textit{all }the UIR of $U(6)$,
belonging to a given UIR of $Sp(12,R)$.

Let us consider the rotational limit \cite{3} of the model defined by the
chain:

\begin{equation}
U(6)\supset SU(3)\times U(2)\supset SO(3)\times U(1)  \label{chain}
\end{equation}%
\begin{equation}
\lbrack N]\ \ \ \ \ \ (\lambda ,\mu )\ \ \ \ \ (N,T)\ \ K\ \ \ \ \ L\ \ \ \
\ \ \ \ \ \ T_{0}  \label{qnum}
\end{equation}%
where the labels below the subgroups are the quantum numbers (\ref{qnum})\
corresponding to their \ irreducible representations. Their values are
obtained by means of standard reduction rules and are given in \cite{3}. In
this limit the operators of the physical observables are the angular
momentum operator 
\[
L_{M}=-\sqrt{2}\sum_{M,\alpha }\ A_{M}^{1}(\alpha ,\alpha ) 
\]%
\bigskip and the truncated (\textquotedblright Elliott\textquotedblright )\
quadrupole operator 
\[
Q_{M}=\sqrt{6}\sum_{M,\alpha }A_{M}^{2}(\alpha ,\alpha ), 
\]%
which define the algebra of $SU(3)$.

The ''pseudospin'' and number of bosons operators: 
\begin{eqnarray*}
T_{+1} &=&\sqrt{\frac{3}{2}}A^{0}(p,n);\ \ \text{\ \ \ \ \ \ \ \ \ \ \ \ \ \
\ \ \ \ \ \ \ \ \ \ \ \ \ \ }\ T_{-1}=-\sqrt{\frac{3}{2}}A^{0}(n,p); \\
T_{0} &=&-\sqrt{\frac{3}{2}}[A^{0}(p,p)-A^{0}(n,n)];\ \ \text{\ \ \ \ \ \ \ }%
\ N=-\sqrt{3}[A^{0}(p,p)+A^{0}(n,n)],
\end{eqnarray*}
define the algebra of $U(2)$.

Since the reduction from $U(6)$ to $SO(3)$ is carried out by the mutually
complementary groups $SU(3)$ and $U(2)$, their quantum numbers are related
in the following way: 
\begin{equation}
T=\frac{\lambda }{2},N=2\mu +\lambda  \label{NTcon}
\end{equation}%
Making use of the latter we can write the basis as 
\begin{equation}
\mid \lbrack N]_{6};(\lambda ,\mu =\frac{N}{2});K,L,M;T_{0}\rangle =\mid
(N,T);K,L,M;T_{0}\rangle  \label{bast}
\end{equation}%
The ground state of the system is: 
\begin{equation}
\mid 0\text{\ \ }\rangle =\mid (0,0);0,0,0;0\rangle =\mid
(N=0,T=0);K=0,L=0,M=0;T_{0}=0\text{ }\rangle
\end{equation}%
which is the vacuum state for the $Sp(12,R)$ group.

\bigskip Then the basis states \cite{3} associated with the even irreducible
representation of the $Sp(12,R)$ can be constructed by the application of
powers of raising generators $F_{M}^{L}(\alpha ,\beta )$ of the same group.
The $SU(3)$ representations $(\lambda ,\mu )$ are symmetric in respect to
the sign of $T_{0}.$

Hence, in the framework of the discussed boson representation of the $%
Sp(12,R)$ algebra all possible irreducible representations of the group $%
SU(3)$ are determined uniquely through all possible sets of the eigenvalues
of the Hermitian operators $N$,$T^{2}$, and $T_{0}.$ The equivalent use of
the $(\lambda ,\mu )$ labels facilitates the final reduction to the $SO(3)$
representations, which define the angular momentum $L$ and its projection $%
M. $ The multiplicity index $K$ appearing in this reduction is related to
the projection of $L$ in the body fixed frame and is used with the parity to
label the different bands in the energy spectra of the nuclei. The parity of
the states is defined as $\pi =(-1)^{T}$. This allows us to describe both
positive and negative bands.

\bigskip The Hamiltonian, corresponding to this limit of IVBM\ is expressed
in terms of the first and second order invariant operators of the different
subgroups in the chain (\ref{chain}): 
\begin{equation}
H=aN+\alpha _{6}K_{6}+\alpha _{3}K_{3}+\alpha _{1}K_{1}+\beta _{3}\pi _{3},
\label{Hl}
\end{equation}%
where $K_{n}$ are the quadratic invariant operators of the $U(n)$ - groups
in (\ref{chain}):, $\pi _{3}$ is \ the $SO(3)$ Casimir operator. As a result
of the connections (\ref{NTcon}) the Casimir operators $K_{3}$ with
eigenvalue $(\lambda ^{2}+\mu ^{2}+\lambda \mu +3\lambda +3\mu ),$ is
express in terms of \ the operators $N$ and $T$: 
\[
K_{3}=2Q_{2}+\frac{3}{4}L^{2}=\frac{1}{2}N^{2}+N+T^{2} 
\]

After some transformations the Hamiltonian (\ref{Hl}) takes the following
form 
\begin{equation}
H=aN+bN^{2}+\alpha _{3}T^{2}+\beta _{3}\pi _{3}+\alpha _{1}T_{0}^{2},
\end{equation}%
and is obviously diagonal in the basis (\ref{bast}) labeled by the quantum
numbers of the subgroups of chosen chain (\ref{chain}). Its eigenvalues are
the energies of the basis states of the boson representations of $Sp(12,R)$: 
\begin{equation}
E((N,T);KLM;T_{0})=aN+bN^{2}+\alpha _{3}T(T+1)+\beta _{3}L(L+1)+\alpha
_{1}T_{0}^{2}  \label{Spectr}
\end{equation}

Using the $(\lambda ,\mu )$ labels facilitates and choosing for instance $%
(\lambda ,0)$ multiplet \ together with the reducing rules (\ref{NTcon})
after simple regrouping of the terms in (\ref{Spectr}) we can write the
energy spectrum corresponding to this $(\lambda ,0)$ multiplet as:

\begin{equation}
E(\lambda )=A\lambda -B\lambda ^{2}+C  \label{Elam}
\end{equation}

here $A$, $B$ and $C$ are the combinations of free model parameters of (\ref%
{Spectr}) $a$, $b$, $\alpha _{3}$,$\beta _{3}$ and $\alpha _{1}$.

Hence choosing any permitted by (\ref{NTcon}) $(\lambda ,\mu )$ multiplet\
we again may classify the low lying excited states energies in even even
nuclei applying the parabolic type distribution function and considering
label $\lambda $ as a measure of collectivity of the corresponding excited
states possessing different from $0$ spins. In\textbf{\ Figure 3}. and 
\textbf{Figure 4}. are shown some examples for the classification (consider
also $n=\frac{\lambda }{4}$) of the energies of $2^{+}$,$4^{+}$ , $6^{+}$%
,and $8^{+}$excited states in $^{162}Dy$ and also $2^{+}$ states in $^{240}Pu
$ and $^{250}Cf$ isotopes.

The experimental energies with great accuracy follow the parabolic
distribution function (\ref{Elam}) and similar agreement can be obtained for
all spectra in even even nuclei. All experimental data are taken from \cite%
{4} .

\ We hope that this new interpretation of the experimental data for low
lying collective excited states of even-even nuclei may be useful for divers
aims in nuclear structure models. We also want to believe that it may be in
help for experimentalists investigating low energies nuclear spectra
especially when any ambiguous definition of the states spins exists.

{\large \bigskip }

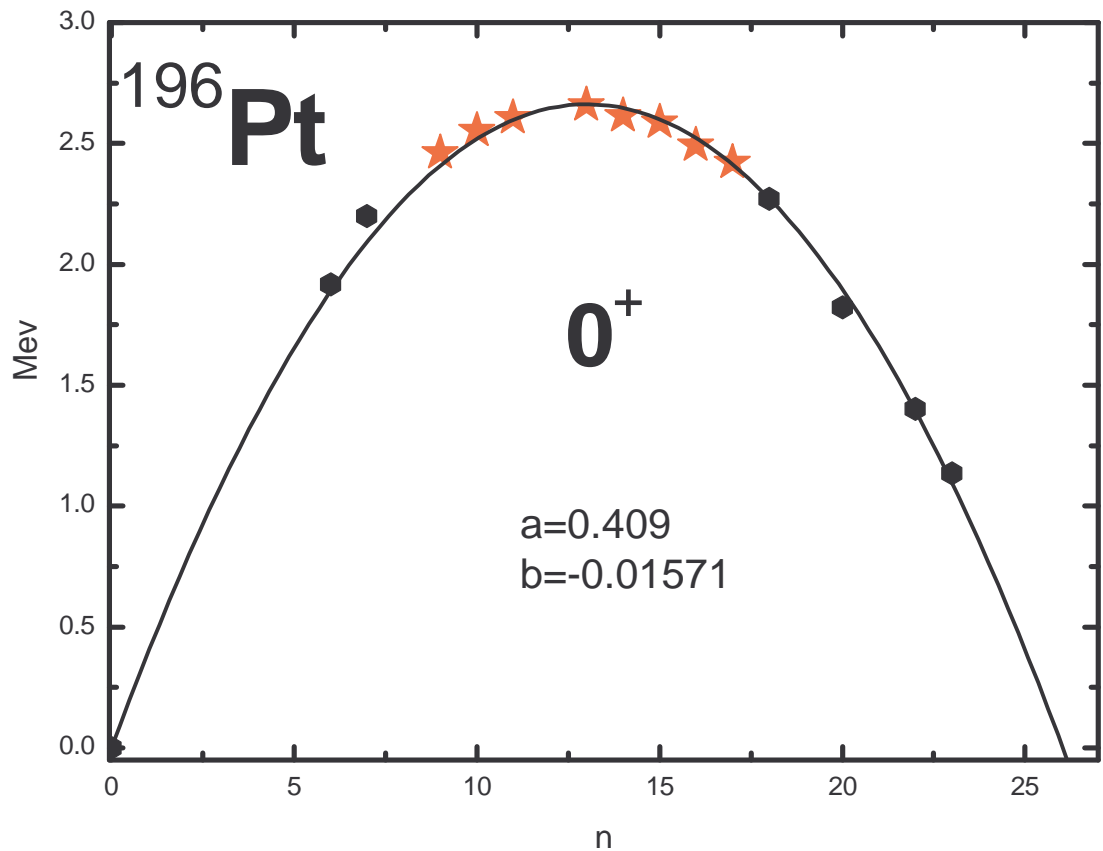
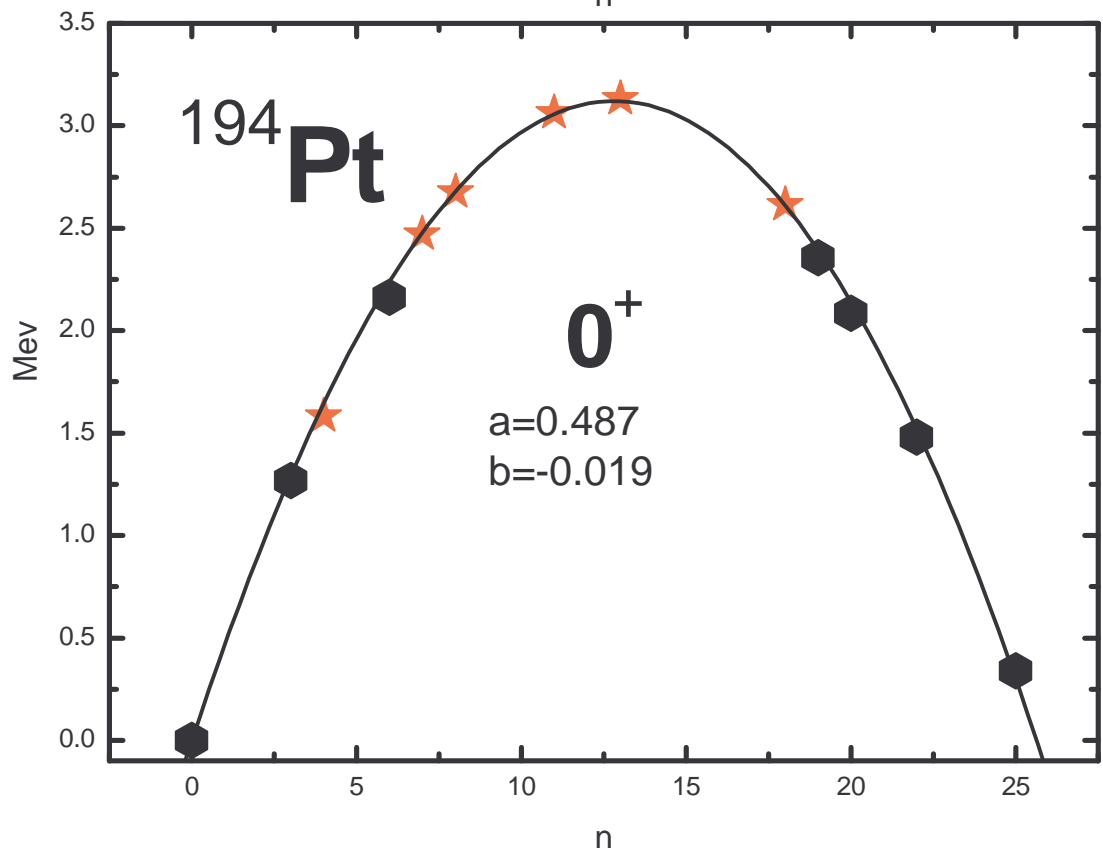
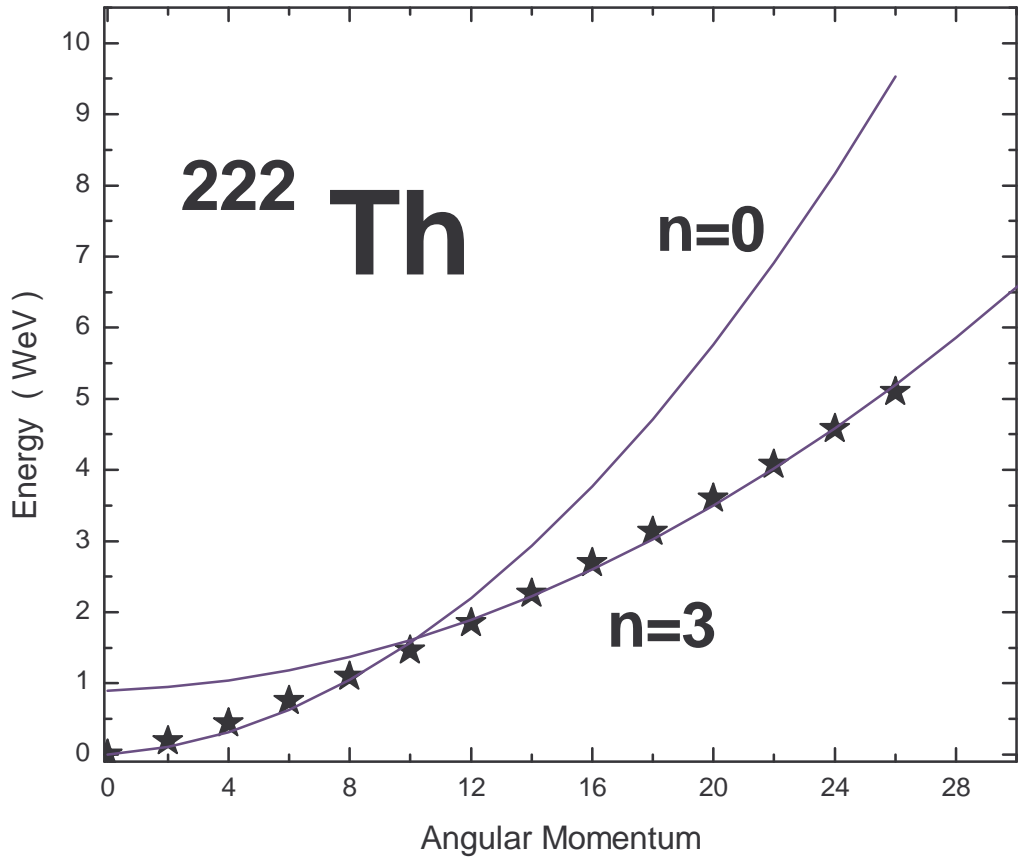
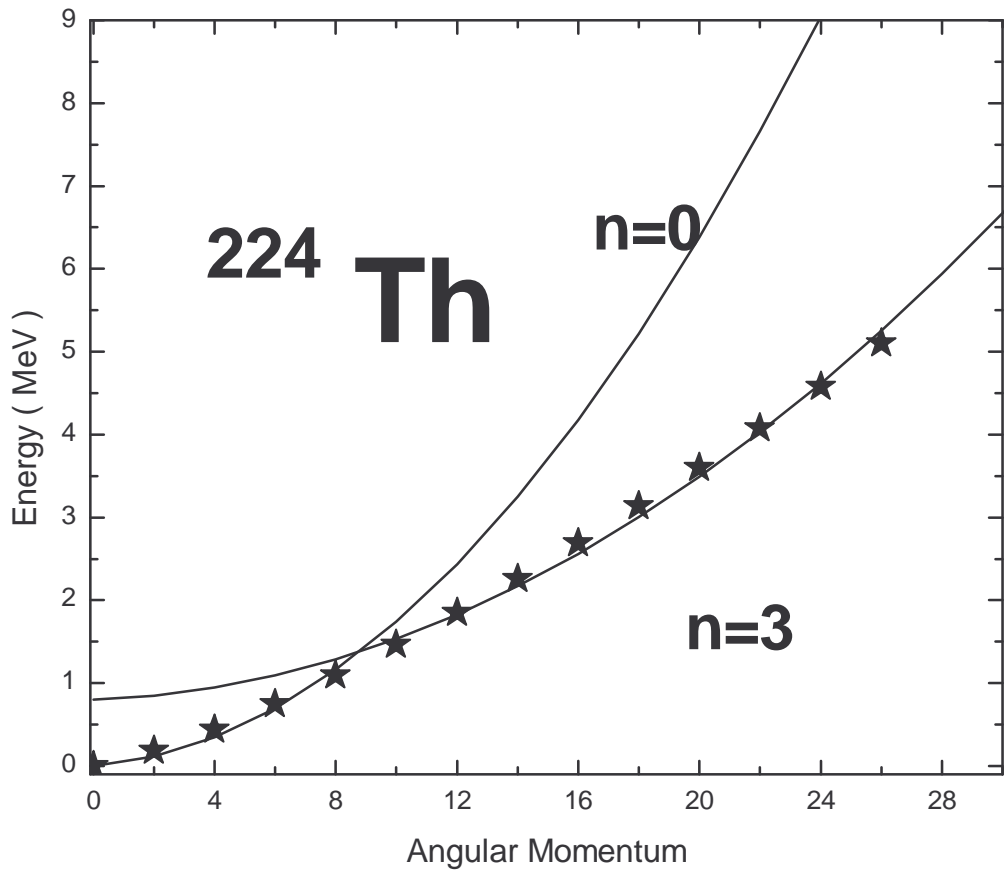
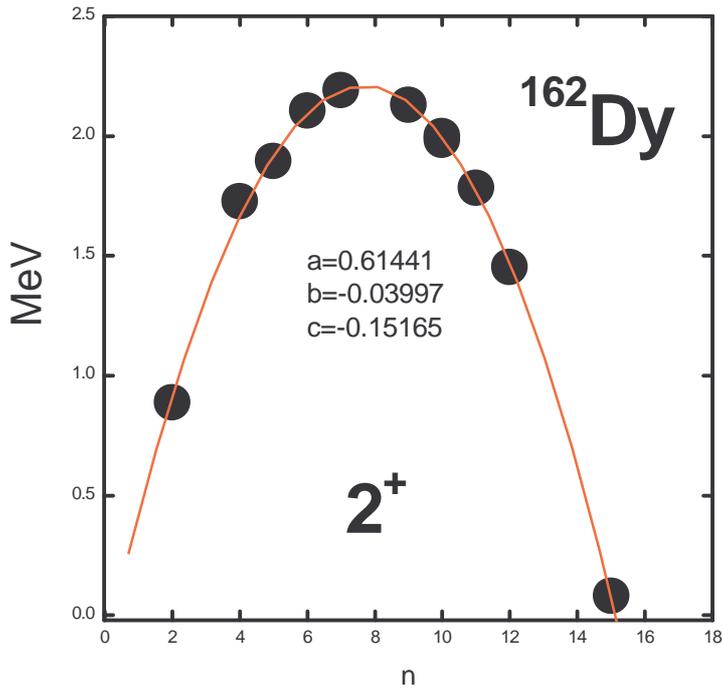
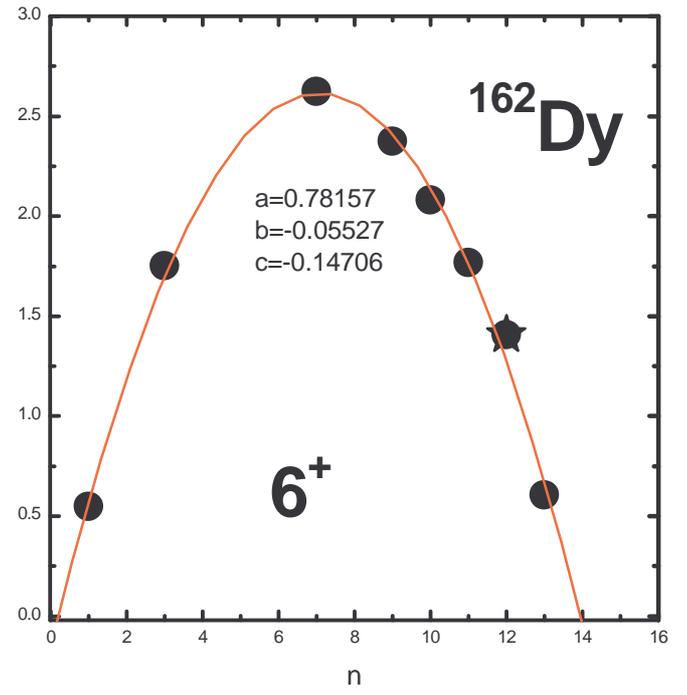
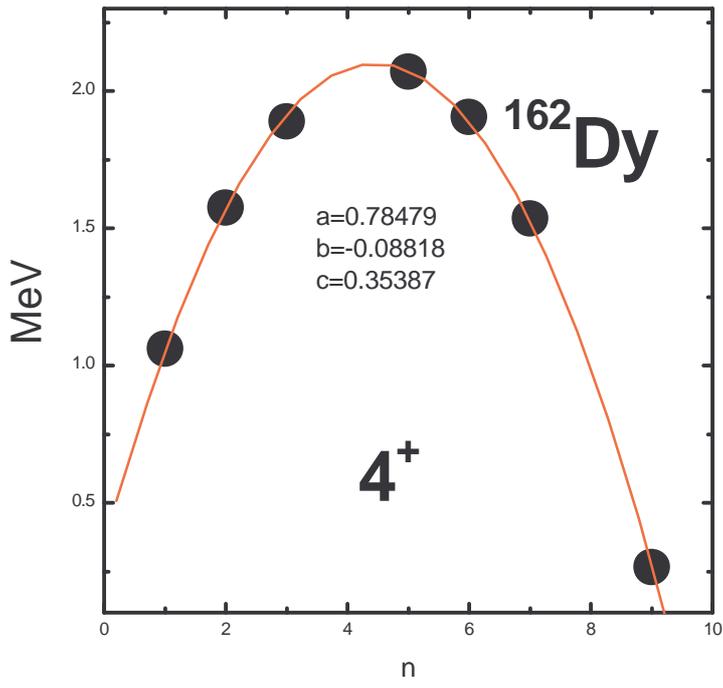
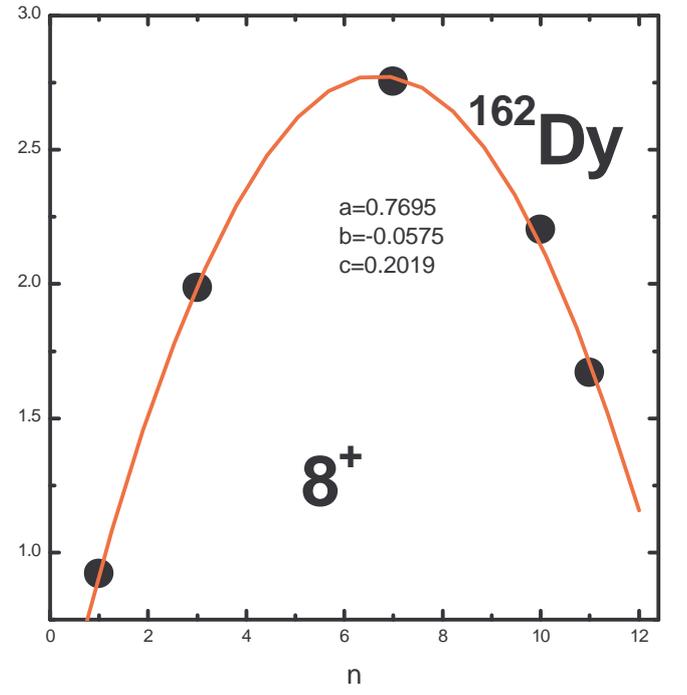
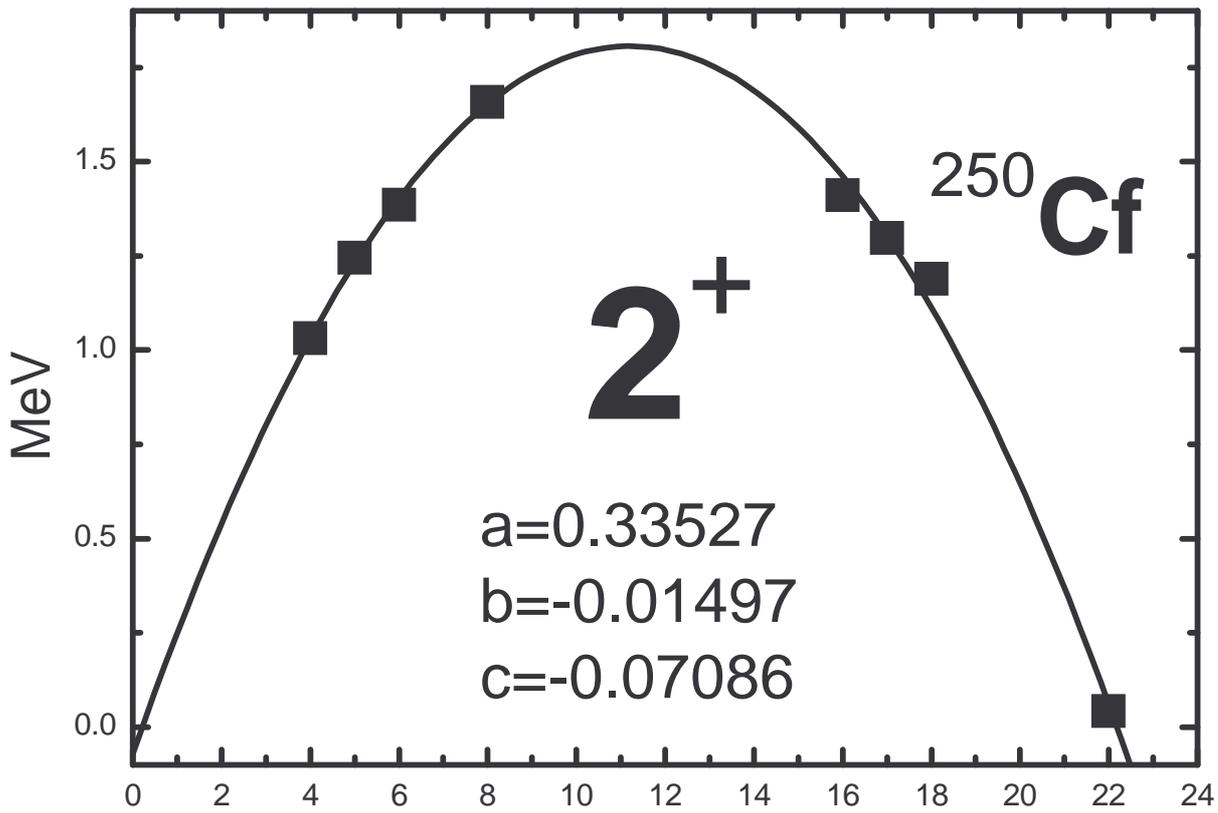
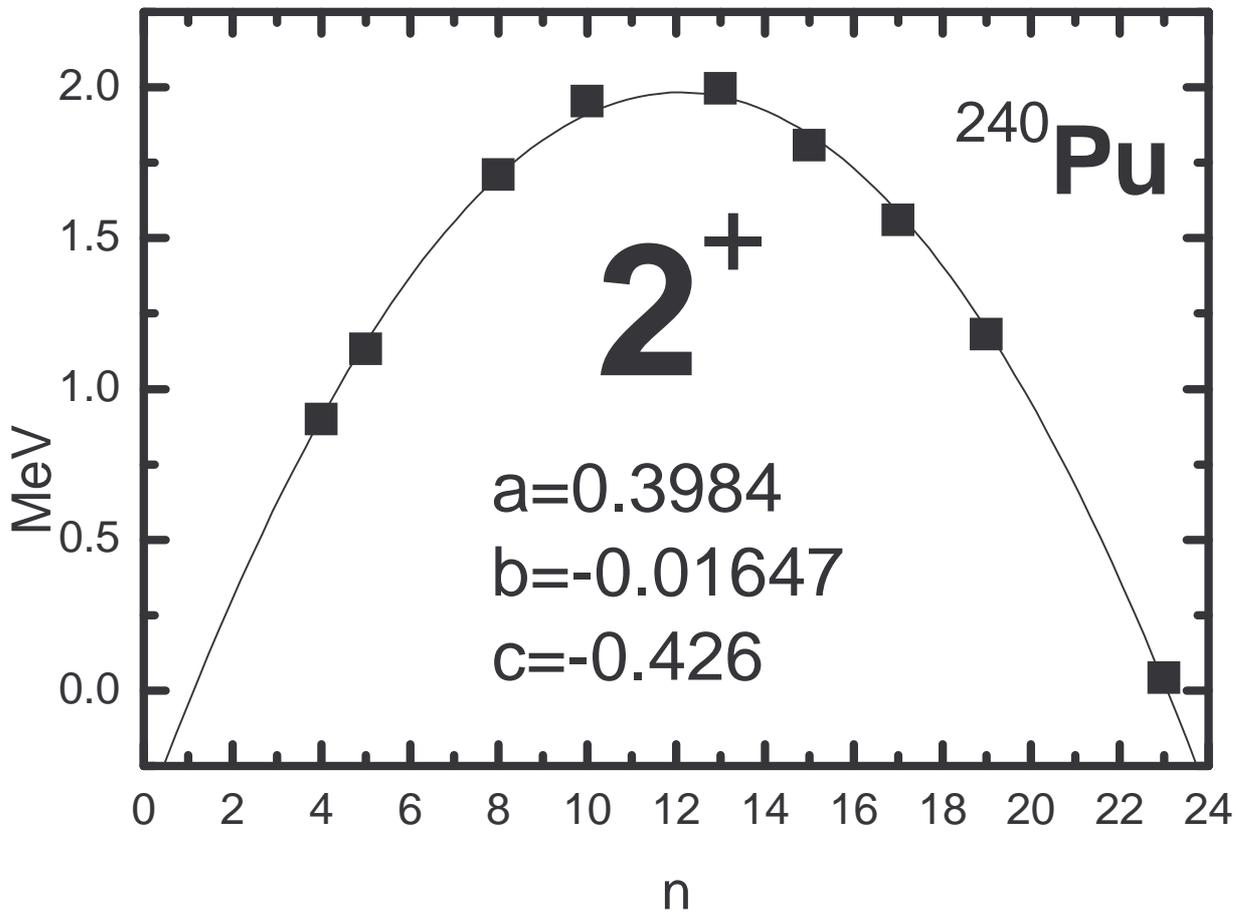
\end{document}